# Edge-guided and Cross-scale Feature Fusion Network for Efficient Multi-contrast MRI Super-Resolution

Zhiyuan Yang[1][0009-0007-8277-5772], Bo Zhang[2][0009−0002−3736−4070], Zhiqiang Zeng[3] and Si Yong Yeo[4][0000−0001−6403−6019]

[1] School of Electronic and Information Engineering, Beihang University, Beijing, China
[2] College of Computing and Data Science, Nanyang Technological University, Singapore
[3] Beijing Institute of Remote Sensing Equipment
[4] Lee Kong Chian School of Medicine, Nanyang Technological University, Singapore
zyyang0416@buaa.edu.cn

**Abstract.** In recent years, MRI super-resolution techniques have achieved great success, especially multi-contrast methods that extract texture information from reference images to guide the super-resolution reconstruction. However, current methods primarily focus on texture similarities at the same scale, neglecting cross-scale similarities that provide comprehensive information. Moreover, the misalignment between features of different scales impedes effective aggregation of information flow. To address the limitations, we propose a novel edge-guided and cross-scale feature fusion network, namely ECFNet. Specifically, we develop a pipeline consisting of the deformable convolution and the cross-attention transformer to align features of different scales. The cross-scale fusion strategy fully integrates the texture information from different scales, significantly enhancing the super-resolution. In addition, a novel structure information collaboration module is developed to guide the super-resolution reconstruction with implicit structure priors. The structure information enables the network to focus on high-frequency components of the image, resulting in sharper details. Extensive experiments on the IXI and BraTS2020 datasets demonstrate that our method achieves state-of-the-art performance compared to other multi-contrast MRI super-resolution methods, and our method is robust in terms of different super-resolution scales. Our code is available at https://github.com/zhiyuan-yang/Edge-Guided-Cross-Scale-MRI-Super-resolution.

## 1 Introduction

Magnetic resonance imaging (MRI) is a non-invasive and radiation-free imaging technique that plays a unique and essential role in clinical diagnosis. Compared to other imaging techniques, it can visualize anatomical tissues of different parts of the human body. Despite its advantages, the acquisition of high-resolution (HR) MRI images faces

---

challenges such as limited scanning time and patient motion [1, 2]. Therefore, MRI super-resolution (SR) has always been an important research topic in the clinical imaging community.

Traditional SR techniques such as interpolation and dictionary learning often result in over-smoothed or blurred images [3]. In recent years, deep learning (DL) based methods [4, 5] have attracted much attention, demonstrating remarkable performance. DL-based methods can be categorized into two types: single-contrast methods and multi-contrast methods. Compared to single-contrast methods, multi-contrast SR methods, which leverage complementary information from different contrasts, have shown to be more powerful. MRI routinely generates multi-contrast images with different acquisition time: T1-weighted (T1W) images normally require shorter scanning time than T2-weighted (T2W) images, so clinicians usually acquire HR T1W images (fully-sampled) and low-resolution (LR) T2W images (under-sampled). They provide complementary information about the anatomical structure, and therefore it is natural to use T1W images as the reference to acquire SR T2W images.

Reference-based SR techniques have been extensively used for both natural images and medical images [6, 7]. TTSR [8] proposes to use the hard attention mechanism to search for the most spatially relevant patch in reference images and integrate it with LR features to generate HR details. Subsequently, MASA [9] and McMRSR [10] develop a coarse-to-fine patch matching scheme that significantly reduces the computation cost and achieves better performance. In addition, WavTrans [11] incorporates the wavelet transformation into the SR framework to capture both high-frequency local structures and global information. However, these patch-based matching methods only utilize the most relevant patch in the reference images, which may overlook the complex relationship between the reference images and the LR images. Moreover, most methods adopt a straightforward approach to transfer the texture information, using either simple feature concatenation [9] or multiplication [8]. To overcome this limitation, the attention mechanism and the fully-powered transformer architecture have been introduced to extract the correlation between the LR features and the reference images. MINet [12] uses a channel-spatial attention module to fuse the features of different stages, while DCAMSR [13] proposes a dual cross-attention transformer to capture the complementary information between multi-contrast images.

Although these recent multi-contrast methods [8-13] have achieved desirable results, there are still some challenges: First, reference-based methods exclusively consider texture transfer from the reference modality, while neglecting the intrinsic anatomical structure. This neglect may lead to superficial and inconsistent SR results. In medical image analysis, it is essential to preserve the anatomical structure in the images for accurate diagnosis. Traditional methods have established the significance of incorporating structure information as a valuable prior constraint [14–17], which allows more attention to be allocated to the image details. Second, current methods only utilize texture similarities of the same scale. Earlier studies [18] have suggested that texture similarities in MRI are not only at the same scale but also across scales. Leveraging features at varying scales to aggregate the information flow can enhance the SR performance. However, simply fusing multi-scale features may introduce more redundant noise due to the misalignment between features of different scales.

To overcome these challenges, we propose ECFNet which adopts several customized modules for SR of biomedical imaging data, depicted in Fig. 1. In particular, we adopt a coarse-to-fine feature fusion strategy to generate texture information. In addition, we add a structure branch containing high-frequency components to assist the model in generating sharper details. The proposed method effectively learns the previously neglected features of different granularities through a multi-scale feature fusion strategy and incorporates the structure information, leading to accurate SR of details. The contributions of this paper are summarized as follows: 1) We introduce the cross-scale feature fusion module (CFFM) that effectively aligns and fuses features of different scales, enhancing the aggregation of information flow. 2) The texture transfer module (TTM) is proposed to adaptively remap the distribution of reference texture with LR features so that the network can better utilize the reference information. 3) We introduce the structure information collaboration module (SICM), which facilitates interaction between features and structure information. The SICM enables the network to allocate more attention to the details while preserving the anatomical structure. Extensive experiments on two public datasets, the IXI [19] and BraTS2020 [20] datasets, demonstrate that our method achieves state-of-the-art performance.

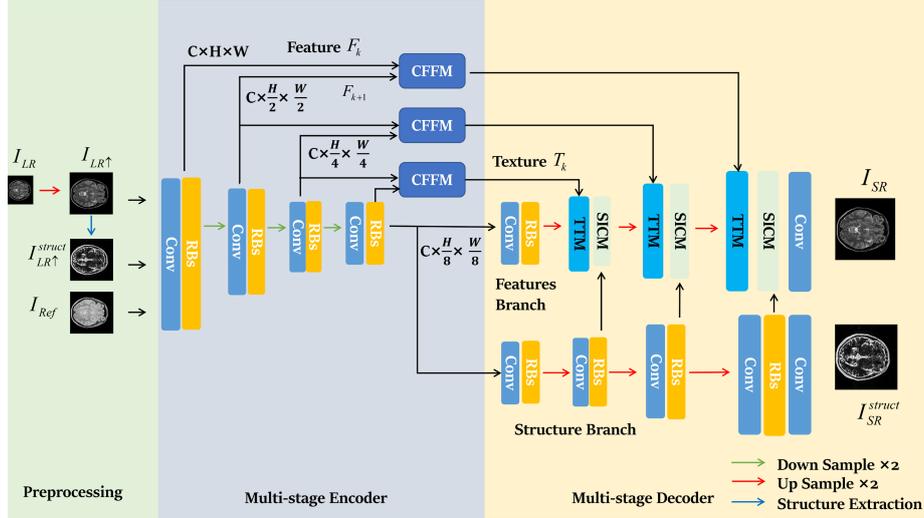

**Fig. 1.** The overall architecture of our proposed ECFNet.

## 2    Methodology

### 2.1    Overall Architecture

Given the LR image $I_{LR}$ (i.e. the T2W images) and the corresponding reference image (Ref) $I_{Ref}$ (i.e. the T1W images), the ECFNet can accurately restore $I_{LR}$ to the SR image $I_{SR}$. As shown in Fig 1, the framework mainly consists of three parts: the preprocessing, the multi-stage encoder, and the multi-stage decoder. In the preprocessing stage, $I_{LR}$ is first interpolated to the same size as $I_{Ref}$, and the Sobel operator is used to extract the edge map $I_{LR\uparrow}^{struct}$. The multi-stage encoder contains four layers, where each layer consists of a down-sample convolutional layer and residual blocks. After passing $I_{LR\uparrow}$ and $I_{Ref}$ into the encoder, features with different scales are obtained, denoted as $F_k$ and $F_k^{Ref}$ where $k=1,2,3,4$. The CFFM aligns and fuses the features extracted from $I_{Ref}$ and $I_{LR\uparrow}$ to generate the coarse-to-fine texture $T_k$. In the multi-stage decoder, the texture is first aggregated with the features using the TTM. After that, the SICM facilitates interaction between the features and structure information, refining the details according to the structure information. Finally, we obtain the SR image $I_{SR}$ and the SR structure map $I_{SR}^{struct}$ using a simple convolutional layer.

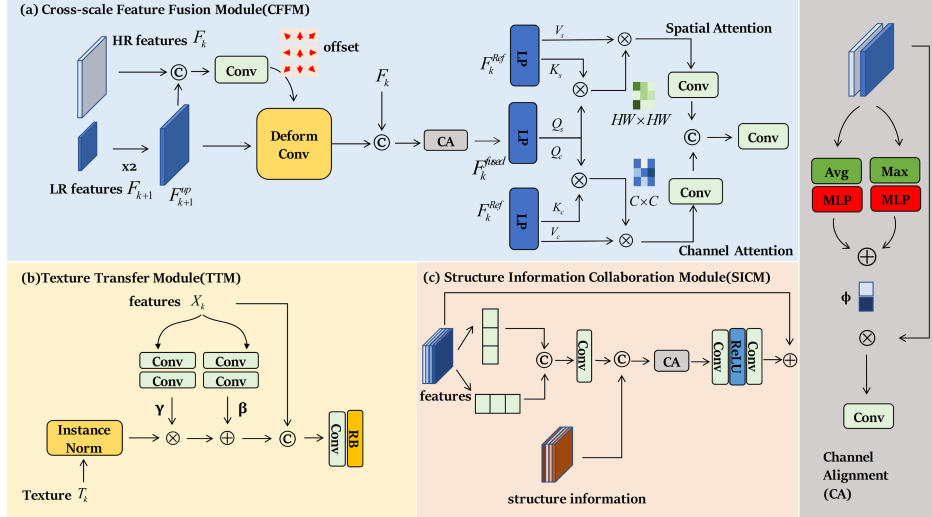

**Fig. 2.** The components of ECFNet: (a) Cross-scale Feature Fusion Module (CFFM); (b) Texture Transfer Module (TTM); (c) Structure Information Collaboration Module (SICM).

## 2.2   Cross-scale Feature Fusion Module

In CFFM shown in Fig. 2 (a), LR features $F_k$ are aligned and then fused with Ref features $F_k^{Ref}$ to incorporate the information from reference images. Since cross-scale similarities of features are widespread, utilizing the aggregated information from different scales can improve the SR results. The key issue is that misalignment of position and channel may appear across different layers, which hinders the comprehensive integration of multi-scale information. Therefore, we propose to adaptively align the up-sampled LR features $F_{k+1}^{up}$ with HR features $F_k$.

First, deformable convolution [21] is used to introduce learnable offsets to the spatial sampling locations, augmenting the alignment of the receptive field in the down-sampled features with HR features. The offset is learned by convolutional layers from concatenated features:

$$offset = Conv_{3\times3}(Concat(F_k, F_{k+1}^{up})). \quad (1)$$

The deformable convolution then uses the offset to get the aligned features:

$$F_{k+1}^{aligned}(p_i) = \sum_{p_n \in \mathcal{R}} w(p_n) \cdot F_{k+1}^{up}(p_i + p_n + \Delta p_n). \quad (2)$$

where $p_i$ denotes a pixel in aligned features $F_{k+1}^{aligned}$, $w(\cdot)$ and $\Delta p_n$ are the weight and the offset respectively. The aligned LR features $F_{k+1}^{aligned}$ are subsequently concatenated with the HR features $F_k$. To alleviate the channel misalignment among features of different scales, we introduce a channel alignment (CA) module. The global max pooling layer and global average pooling layer are used to extract the channel information from the concatenated features $F_k^{concat}$ respectively, and the outcome is denoted as $P_{avg} \in \mathbf{R}^{C\times1\times1}$ and $P_{max} \in \mathbf{R}^{C\times1\times1}$. A multi-layer perceptron (MLP) consisting of two fully connected layers with a reduction rate of 16 is then used to get the channel alignment coefficient $\phi$. The output is obtained by:

$$F_k^{fused} = \phi \cdot F_k^{concat} + F_k^{concat}. \quad (3)$$

After aggregating the features at different scales, a dual cross-attention transformer [13] is used to generate reference texture by utilizing the complementary information from $F_k^{Ref}$. Linear projection functions are used to compute the query, value, and key of the features, and the spatial and channel attention are then obtained by

$$T_s = softmax(\frac{Q_s \times K_s^T}{\sqrt{d}} \times V_s), \quad (4)$$

$$T_c = softmax(\frac{Q_c \times K_c^T}{\sqrt{d}} \times V_c). \quad (5)$$

Finally, the spatial and channel attention are concatenated and reduced to half channel with depth-wise convolution. The obtained features are then processed by the residual blocks to generate textures $T_k$.

### 2.3    Multi-stage Decoder

At each stage of the decoder, the extracted texture is first integrated with the features using TTM, where the distribution of the texture is remapped with the features. The details are then refined according to the structure information using the SICM so that more attention is allocated to them.

Since the distribution of the extracted texture may be inconsistent with LR features, simple concatenation may lead to suboptimal results. Inspired by the work of [9], we design a texture transfer module (TTM) as shown in Fig 2 (b) to remap the distribution of the texture with LR features. The instance normalization is used to extract the structure of texture and discard its style:

$$T_k \leftarrow \frac{T_k - \mu_{T_k}}{\sigma_{T_k}}. \tag{6}$$

After that, the affine transformation is used to update the features:

$$T_k \leftarrow X_k \otimes \beta + \gamma. \tag{7}$$

Two separate convolutional blocks are used to learn the affine transformation parameters $\beta$ and $\gamma$ so that the features can adapt the style to the texture while maintaining the structure. Then the transferred texture is concatenated with the features and fused by a residual block. Compared to simple multiplication or concatenation, the TTM takes characteristics of both features and texture into consideration. The adaptive fusion process can enhance the incorporation of reference information.

MRI has a large plain background and small important target areas. These areas contain rich tissue information that is important for accurate diagnosis. The edge map corresponds to the high-frequency components in the images, therefore incorporating the edge information can guide the network to allocate more attention to the details during the SR reconstruction. Since the edge map has zero values in most areas, an asymmetric convolutional group consisting of 1×3 and 3×1 convolutions is used to extract geometric structure both vertically and horizontally, and 1×1 convolution is adopted to refine the features:

$$X_k^{edge} = Conv_{1\times1}(Concat[Conv_{3\times1}(X_k), Conv_{1\times3}(X_k)]). \tag{8}$$

The channel alignment (CA) module is adopted to remap the distribution of structure information with the features. Next, to improve the stability of the network training, we use the residual connection to get the fused features:

$$X_{k-1} = Conv(\text{ReLU}(Conv(X_k^{aligned}))) + X_k. \tag{9}$$

The SICM makes the network easier to preserve the anatomical information for accurate SR and leads to sharper details.

### 2.4   Loss Function

The $L_1$ loss is used for the reconstruction and structure loss. The total loss function is

$$L = \frac{1}{N} \sum_{n=1}^{N} L_1(I_{SR}, I_{HR}) + L_1(S(I_{HR}), I_{SR}^{struct}), \tag{10}$$

where $S(\cdot)$ represents the Sobel operator.

## 3    Experiments

**Datasets and Baselines**. The IXI [19] and BraTS2020 [20] datasets are used to evaluate our proposed method. The IXI dataset contains registered T2W and proton density weighted (PDW) 3D MRI volumes of 578 subjects, and we use the PDW MRI as the reference modality. We adopt the same preprocessing procedure of [22], where 3D volumes are clipped into the size of $240 \times 240 \times 96$. 500 subjects are randomly selected as the training set and another 70 subjects as the testing set. For each subject, 10 slices are selected. 2-fold and 4-fold down-sampled T2W LR images are created using the k-space truncation. The BraTS2020 dataset contains 369 subjects for the training dataset and 125 subjects for the validation dataset. Each subject has 4 modalities with size of $240 \times 240 \times 155$, and we use T1W images as reference images. 300 subjects are randomly chosen for training and another 100 subjects for testing. We compare our methods with four multi-contrast methods (MINet [12], DCAMSR [13], TTSR [8], WavTrans [11]) and a single-contrast method (SwinIR [23]). Peak signal-to-noise ratio (PSNR) and structure similarity index measure (SSIM) are used to evaluate the performance of different methods.

**Table 1.** Quantitative results on two datasets with different scales. Red numbers indicate the best result, and blue numbers indicate the second-best result.

| Dataset | IXI | | | | BraTS2020 | | | |
|---|---|---|---|---|---|---|---|---|
| Scale | 2× | | 4× | | 2× | | 4× | |
| Metric | PSNR | SSIM | PSNR | SSIM | PSNR | SSIM | PSNR | SSIM |
| TTSR | 38.487 | 0.981 | 30.400 | 0.920 | 39.597 | 0.990 | 32.671 | 0.961 |
| SwinIR | 37.002 | 0.977 | 29.250 | 0.908 | 39.418 | 0.991 | 31.758 | 0.956 |
| MINet | 39.925 | 0.984 | 34.093 | 0.942 | 40.315 | 0.992 | 33.602 | 0.965 |
| DCAMSR | 40.324 | 0.986 | 35.908 | 0.967 | 40.175 | 0.991 | 33.806 | 0.967 |
| WavTrans | 39.719 | 0.981 | 33.443 | 0.963 | 39.857 | 0.995 | 33.406 | 0.971 |
| Ours | 41.823 | 0.987 | 37.213 | 0.970 | 41.218 | 0.995 | 34.985 | 0.972 |

**Implementation Details**. We train all the models on NVIDIA GeForce RTX 3090 GPUs. Our model is trained using the Adam optimizer with a learning rate of 2e-4 for 50 epochs. The batch size is set as 10. The parameters of the Adam optimizer, $\alpha$ and $\beta$, are set to 0.9 and 0.999 respectively. All the compared models are trained using their default parameter settings.

8     Z. Yang et al.

Table 2. Ablation study on the IXI dataset with 4-fold SR.

| Variant | Modules | | | Metrics | |
|---|---|---|---|---|---|
| | CFFM | TTM | SICM | PSNR | SSIM |
| *w/o* multi-scale feature alignment | × | √ | √ | 33.478 | 0.941 |
| *w/o* texture transfer | √ | × | √ | 35.437 | 0.968 |
| *w/o* structure branch | √ | √ | × | 35.312 | 0.965 |
| full version | √ | √ | √ | 37.213 | 0.970 |

**Quantitative Results**. Table 1 summarizes the PSNR and SSIM scores on two public datasets in 2-fold and 4-fold SR. Compared with other methods, our method achieves the best results in all cases, which proves the effectiveness of our method. In the challenging 4-fold SR situation, our method can still achieve a desirable result. We give the multi-feature fusion strategy credit for it. It aggregates information flow from different scales so that even in the LR situation, the network is still able to extract effective texture information. Besides, the alignment procedure can effectively reduce the redundant noise when fusing different scale features.

**Qualitative Results.** Fig. 3 shows the SR results and the corresponding error maps on two datasets with different SR rates. In the error maps, prominent features indicate poor detail reconstruction. It can be observed that our method is superior compared to other methods in both datasets, which proves the robustness of our method. Furthermore, our method generates sharper texture details compared to other methods because we incorporate the structure information, allowing the network to focus on the details during the reconstruction process. In the SR process, the features are adaptively adjusted in the informative regions guided by the structure information, resulting in more details.

**Ablation Study.** We conduct ablation studies on the IXI dataset for 4-fold SR to evaluate the effectiveness of different modules within our framework, and the results are shown in Table 2. Three variant networks are used: 1) *w/o* multi-scale feature alignment, where the cross-scale alignment part in the CFFM is not used. 2) *w/o* texture transfer, which is our model without the TTM. 3) *w/o* structure branch, which is our model without the edge map extraction and the edge branch. The results indicate that the variant *w/o* multi-scale feature alignment performs worst, which proves that our alignment module effectively integrates features from different scales. The degradation of variant *w/o* struct branch is consistent with our conclusion that structure information can enhance the SR reconstruction and lead to sharper details. Furthermore, the improvement from the variant *w/o* TTM to the full version also proves the effectiveness of the texture transfer module.

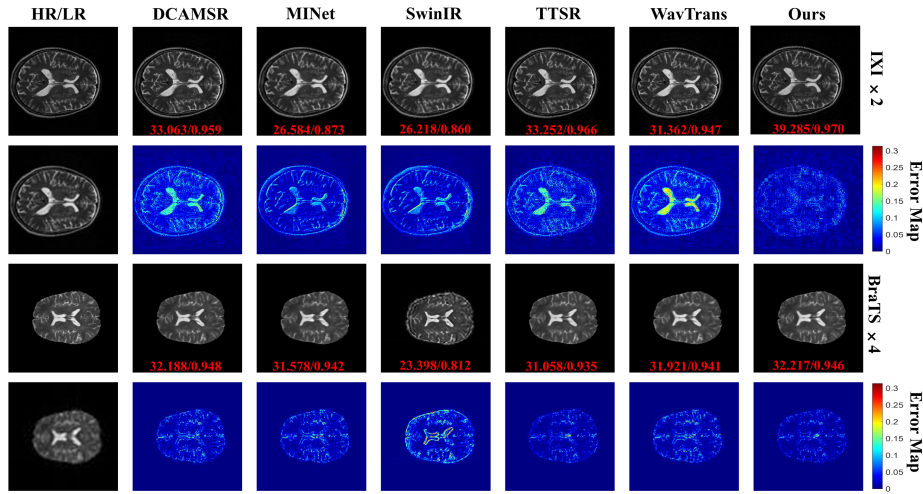

**Fig. 3.** Qualitative results and error maps of different methods on two datasets. The first/third row are the SR results, and the second/fourth row are the corresponding error maps. The brighter color suggests more errors.

## 4     Conclusion

In this study, we propose an edge-guided and cross-scale feature fusion network for multi-contrast MRI super-resolution. Specifically, we design a novel pipeline to utilize cross-scale similarities in MRI that can provide comprehensive information. In addition, we incorporate the structure information to guide the network towards generating sharper textures. Extensive experiments demonstrate that the proposed method achieves state-of-the-art performance, especially in the challenging four-fold SR. Our work provides a possible direction for further research in processing multi-contrast MRI, which has great potential uses in many medical applications. In the future, we would like to explore multi-contrast MRI super-resolution at arbitrary scales.

**Acknowledgments.** This project is supported by the Lee Kong Chian School of Medicine - Ministry of Education Start-Up Grant

**Disclosure of Interests.** Authors have no conflict of interest to declare.